\documentstyle[12pt,epsf]{article}
\def\fun#1#2{\lower3.6pt\vbox{\baselineskip0pt\lineskip.9pt
        \ialign{$\mathsurround=0pt#1\hfill##\hfil$\crcr#2\crcr\sim\crcr}}}

\begin{document}

\title{\vskip-2.5truecm{\hfill \baselineskip 14pt {{
\small  \\       \hfill SISSA--66/96/EP \\ \hfill IC/96/73 \\
\hfill May 1996}}\vskip .1truecm}
 {\bf CP violating decays in \\ leptogenesis scenarios}}

\vspace{2cm}

\author{Laura Covi$^{1}$, Esteban Roulet$^{1}$ and Francesco
Vissani$^2$ \\  \  \\
 $^{1}${\it International School for Advanced Studies, 
SISSA-ISAS},\\ {\it Via Beirut 2/4, I-34014, Trieste, Italy} \\ $^{2}$
{\it International Centre for Theoretical Physics, ICTP,}\\{\it 
Strada Costiera 11, I-34100, Trieste, Italy}}

\date{}
\maketitle
\vfill

\begin{abstract}
\baselineskip 14pt

We compute the CP violation in the decays of heavy electroweak singlet
neutrinos, arising from both the one--loop vertex corrections and 
the wave function mixing. We extend the computation to the
supersymmetric version of the model and discuss the implications for
the generation of a lepton number asymmetry by the out of equilibrium
decay of the heavy (s)neutrinos in the early Universe, to be
reprocessed later in the observed baryon excess by anomalous
electroweak processes. 

\end{abstract}
\vfill
\thispagestyle{empty}

\newpage
\pagestyle{plain}
\setcounter{page}{1}
\def\beq{\begin{equation}}
\def\eeq{\end{equation}}
\def\beqa{\begin{eqnarray}}
\def\eeqa{\end{eqnarray}}
\baselineskip 20pt
\parskip 8pt
 
A very attractive scenario, proposed by Fukugita and Yanagida
\cite{fu86}, for the generation of the baryon asymmetry of the
Universe, is based on the production of a lepton asymmetry by the out
of equilibrium decays of heavy ($m\gg{\rm TeV}$) electroweak singlet
neutrinos. Their decays into light leptons and Higgs bosons can
violate CP if the Yukawa couplings involved have unremovable phases,
and can then lead to the production of an excess of antileptons over
leptons in the final state. The lepton asymmetry so produced is then
partially converted into a baryon asymmetry by anomalous electroweak
processes \cite{sphal}, which are in equilibrium at temperatures 
larger than the
electroweak phase transition one, and as a result,
 the amount of baryons present in the Universe, $n_B/n_\gamma\simeq
5\times 10^{-10}$, can be accounted for \cite{lu92,pl96}.

To postulate the existence of right--handed `sterile' neutrinos
constitutes one of the simplest and most economical extensions of the
standard model, being strongly motivated by Grand Unified theories (GUT)
such as SO(10). It also allows to implement the see--saw mechanism which
naturally accounts for non--zero, but still very small, light neutrino
masses. Moreover, the values of the light neutrino masses suggested by
the MSW solution to the solar neutrino problem, the atmospheric
neutrino anomaly and the hot dark matter scenarios \cite{ge95}, 
point towards an
intermediate scale, $M\sim 10^{9}$--$10^{13}$~GeV, for the
right--handed neutrino masses. This further encourages the
consideration of scenarios where the heavy neutrino decays act as
sources for the generation of the baryon asymmetry of the Universe,
since for these large masses it becomes easier to achieve the out of
equilibrium conditions required for the efficient generation of an
asymmetry,  
and also because temperatures comparable to the intermediate scale are
more likely to be produced as a result of the reheating process at the
end of inflation than the temperatures required in the conventional
baryogenesis GUT scenarios.

In addition to the non--supersymmetric version originally considered
by Fukugita and Yanagida, the extension of this scenario to the
supersymmetric case has been studied by Campbell, Davidson and Olive
\cite{ca93}, and also a related scenario where the asymmetry is
produced in the decay of heavy scalar neutrinos produced
non--thermally by the coherent oscillations of the scalar field at
the end of inflation has been discussed by Murayama et
al. \cite{mu93}. A key ingredient for all these scenarios is the
one--loop CP violating asymmetry involved in the heavy (s)neutrino
decay, and the reconsideration of this quantity will be the main issue
of this work. As will be discussed, there are contributions to the
asymmetry that have not been included previously, and we also want to
discuss the different results  present in the literature on this subject.

Let us start with the non--supersymmetric version of the model, with a
Lagrangian given by

\beq
{\cal L}=-\lambda_{ij}\epsilon_{\alpha\beta}\overline{N_j}P_L\ell_i^\alpha
H^\beta+h.c.\ ,
\eeq
where $\ell_i^T=(\nu_i\ l^-_i)$ and $H^T=(H^+\ H^0)$ are the lepton and
Higgs doublets
($\epsilon_{\alpha\beta}=-\epsilon_{\beta\alpha}$, with
$\epsilon_{12}=+1$), and we are taking the $N_j$ to be the Majorana
mass eigenstate fields, of mass $M_j$. 
Since the scale of the heavy neutrinos is so much larger than the
electroweak scale, it is reasonable to work directly in the
 symmetric phase where all particles except the heavy Majorana
neutrinos $N_j$ are massless, and all components of 
the neutral and charged complex Higgs fields
are physical.

The basic quantity we are interested in is
\beq
\epsilon^N_\ell\equiv{\Gamma_{N\ell}-\Gamma_{N\bar\ell}
\over \Gamma_{N\ell}+ \Gamma_{N\bar\ell}},
\eeq
where $\Gamma_{N\ell}\equiv \sum_{\alpha,\beta}\Gamma(N\to \ell^\alpha
H^\beta)$ and  
$\Gamma_{N\bar\ell}\equiv \sum_{\alpha,\beta}\Gamma(N\to
\bar\ell^\alpha H^{\beta \dagger})$ are the $N$
decay rates (in the $N$ rest frame), summed  over the
neutral and charged leptons (and Higgs fields) which appear as final states
in the $N$ decays\footnote{For the Majorana light neutrinos, one
should think of $\ell$ and $\bar\ell$ as being the left and
right--handed helicities of $\nu$.}.
Note that at tree level 
\beq
\Gamma_{N_i\ell}=\Gamma_{N_i\bar\ell}={(\lambda^\dagger\lambda)_{ii}
\over 16\pi}M_i .
\eeq

The asymmetry $\epsilon^N_\ell$ arises from the interference of the
one--loop diagrams depicted in Fig.~1 with the tree level
coupling. The vertex correction in Fig.~1a is the one that is
usually considered, but it has however been pointed out 
\cite{ig79,bo91,li93} 
that the wave function piece in Fig.~1b also contributes to
the asymmetry, in an amount which is typically  comparable to the
vertex contribution. We have computed these CP violating
asymmetries  and we obtain for the vertex contribution to
$\epsilon^N_\ell$ 

\begin{figure}
\centerline{\epsfbox{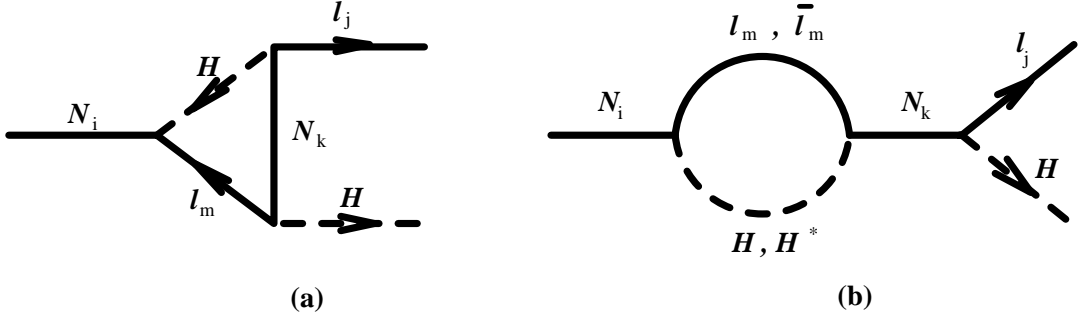}}
\caption{Diagrams contributing to the vertex (Fig.~1a)
and wave function (Fig.~1b) CP violation in the heavy singlet neutrino
decay.}
\end{figure}

\beq
\epsilon^{N_i}_\ell(vertex)={1\over 8\pi}\sum_k f\left(y_k\right)
{\cal I}_{ki} \ ,
\eeq
where $y_k\equiv {M_k^2/M_i^2}$, 
$f(x)=\sqrt{x}\left( 1-(1+x){\rm
ln}\left[(1+x)/x\right]\right)$, and we have defined
\beq
{\cal I}_{ki}\equiv{{\rm Im}\left[
\left(\lambda^\dagger \lambda\right)_{ki}^2\right]\over 
\left(\lambda^\dagger \lambda\right)_{ii}}.
\eeq

For the wave function piece we find
\beq
\epsilon^{N_i}_{\ell_j}(wave)=-{1\over 8\pi}\sum_{k\neq i}{M_i\over
M_k^2-M_i^2}  {{\rm Im}\left\{\left[ M_k
\left(\lambda^\dagger \lambda\right)_{ki}+ M_i
\left(\lambda^\dagger \lambda\right)_{ik}
\right]\lambda^*_{jk}\lambda_{ji}\right\}\over 
\left(\lambda^\dagger \lambda\right)_{ii}}.
\eeq

A useful check of the result can be obtained using the unitarity
relation 
$$
|T_{fi}|^2-|T_{if}|^2\simeq -2\ {\rm
Im}\left[\sum_nT_{ni}T^*_{nf}T_{if}\right], 
$$
where we are retaining only the leading order terms; 
$i$ and $f$ are the initial and final states, and $n$ are the
possible intermediate states connecting them. $T_{fn}$ are the
transition amplitudes between the intermediate and final states, which
in the present case arise from both the $s$ and $t$--channel $N_k$
exchanges, which correspond to wave and vertex contributions
respectively.

The result in Eq.~(4) coincides with the one recently obtained in
ref.~\cite{pl96}, differing by factors of $2$ and $4$ from those in
ref.~\cite{fu94} and \cite{lu92} respectively\footnote{The 
normalization of the $\epsilon$ parameters are sometimes different
(and  not always explicit).}. 
The wave function piece, which in this scenario 
was only considered previously
by Liu and Segr\`e \cite{li93}, is a factor of two larger than their
result due to the fact that their
computation applies to the case in which  the scalar field is real 
and only the neutral lepton runs in the loop. 
Adding the charged lepton loop contribution the result is doubled. 
In Eq.~(6)  we do not
include in the sum over the flavour of the intermediate neutrinos 
the case $k=i$, since in general states degenerate with
the initial one do not contribute to the CP violating asymmetry
\cite{li94}. We also assumed that $|M_k-M_i|\gg \Gamma_{N_k}$ in the
computation. Notice that the contribution from $k=i$ in the
vertex piece, being proportional to
Im$\left[(\lambda^\dagger\lambda)_{ii}^2\right]=0$, also vanishes.
 Hence, both sums in
Eqs.~(4) and (6) may be restricted to $k\neq i$. 
In Eq.~(6) we have not summed over the final lepton flavour, but
after summing over it one gets
\beq
\epsilon^{N_i}_\ell(wave)=-{1\over 8\pi}\sum_{k\neq
i}{M_iM_k\over M_k^2-M_i^2}\ {\cal I}_{ki},
\eeq
so that the piece proportional to $M_i$ in the square brackets in
Eq.~(6) actually gives a null contribution to the total lepton
number asymmetry (although it may generate an asymmetry  in the
individual leptonic numbers $L_j$). In ref.~\cite{li93} it was pointed
out that there should also be 
 in the vertex contribution, in addition to the result in
Eq.~(4), a term like the one just discussed appearing in Eq.~(6). 
This term appears however only when a
single real scalar Higgs field is allowed to run in the loop, but it
can be shown that the new contributions  arising
from the two real scalars in the decomposition of the standard model
Higgs fields $H^\alpha=(H_1^\alpha+iH_2^\alpha)/\sqrt{2}$ 
actually cancel each other, leaving only
the term proportional to ${\cal I}_{ki}$ present in Eq.~(4).

In the particular case in which a hierarchy among the right--handed
neutrino masses  is considered, i.e. for $y_k\gg 1$,
we have that the wave function contribution becomes twice as large as
the  vertex one, and hence 
the total asymmetry produced in the decay of the lightest heavy
neutrino\footnote{In the case of large hierarchies, the asymmetries
produced by the heavier neutrino decays are usually erased before the
lightest one decays.}, $N_1$, becomes
\beq
\epsilon^{N_1}_\ell\equiv \epsilon^{N_1}_\ell(vertex)+
\epsilon^{N_1}_\ell(wave)\simeq
-{3\over 16\pi}\sum_{k\neq 1} {1\over \sqrt{y_k}}\ {\cal I}_{k1}.
\eeq
The inclusion of the wave function piece then increases by a factor of
three  the CP violating
asymmetry  in the
case of large mass hierarchies \cite{li93}.

If the departure from equilibrium in $N_1$ decay is
large, the lepton number asymmetry produced, per unit entropy, is
\beq
{n_L\over s}\simeq {\epsilon^{N_1}_\ell\over s} {g_N T^3\over \pi^2},
\eeq
where $g_N=2$ are the spin degrees of freedom of the Majorana neutrino
$N_1$,  so that their number density (assuming
Maxwell--Boltzmann statistics)  before they
decay is $g_NT^3/\pi^2$ if we assume that they were in chemical 
equilibrium before becoming
non--relativistic (see below). Using that $s=(2/45)g_* \pi^2T^3$,
where $g_*=g_{bosons}+(7/8)g_{fermions}$ is the effective number of
relativistic degrees of freedom contributing to the entropy, ($g_*=106.75$
in the standard non--supersymmetric version of the model), we get
\beq
{n_L\over s}\simeq 4\times 10^{-3}\epsilon^{N_1}_\ell.
\eeq
This lepton asymmetry will then be reprocessed by anomalous
electroweak processes, leading to a baryon asymmetry \cite{ha90}
\beq
n_B= -\left({24+4 N_H\over 66+13 N_H}\right) n_L\simeq -\frac{28}{79} n_L.
\eeq
Here $N_H$ is the number of Higgs doublets ($N_H=1$ in the standard
scenario considered now, 
while $N_H=2$ in the supersymmetric version to be discussed
below if the scalar Higgs bosons  are assumed to be lighter than the
electroweak symmetry breaking scale). 
Combining Eqs.~(10) and (11) and assuming that the Universe expansion is
adiabatic, the present baryon asymmetry which results is
\beq
\frac{n_B}{s}\simeq -1.5\times 10^{-3} \epsilon^{N_1}_\ell.
\eeq
Values of $\epsilon^N_\ell\simeq -5\times 10^{-8}$ are then required to
account for the value $n_B/s\simeq 0.6$--$1\times 10^{-10}$ inferred
from the successful theory of primordial nucleosynthesis \cite{co95}.
In the case in which there is a hierarchy in the heavy neutrino
masses, the Yukawa parameters need then to satisfy
\beq
\sum_{k\neq 1}{M_1\over M_k}\ {\cal I}_{k1}\simeq 0.9\times 10^{-6}.
\eeq

Let us also briefly mention that the lepton asymmetry produced by $N$
decays is smaller than the value in Eq.~(9) if the departure from
equilibrium is not large during the decay epoch, as is the case if the decay
rate $\Gamma_1\equiv\Gamma_{N_1\ell}+\Gamma_{N_1\bar\ell}$ 
is comparable or larger than the expansion rate
of the Universe when $N_1$ becomes non relativistic, i.e. $\Gamma_1\geq
H(T=M_1)$, with $H(T)$ being Hubble's constant at temperature
$T$. We also want to emphasize that a problem of these scenarios is
that the Yukawa interactions are not effective in establishing an
equilibrium population of heavy neutrinos. To see this, 
note that before $N_1$ becomes non--relativistic,
i.e. for $x\equiv M_1/T\ll1$,  the thermally
averaged rate scales as $T^{-1}$, since $\langle
\Gamma_1\rangle=K_1(x)/K_2(x)\Gamma_1\simeq(x/2)\Gamma_1\propto
M_1^2/T$, 
where $K_{1,2}$ are Bessel functions \cite{lu92}.
Taking into account that 
$H\propto T^2/M_{Planck}$ and supposing  that $\Gamma_1\ll H(T=M_1)$ (so that 
the departure from equilibrium during the decay is
significant),   one concludes that $\langle\Gamma_1\rangle\ll H$ holds
true also for all temperatures larger than $M_1$. Hence, 
the inverse decays are
unable to establish chemical equilibrium for $N_1$ 
at any temperature $T>M_1$. The
pair production of $N_1$ from Higgs boson or light lepton scattering may
do somewhat better than inverse decays at high temperatures, since the
relevant rate scales as $\langle \sigma v\rangle\propto T$, but it is
anyhow insufficient to achieve chemical equilibrium for $N_1$ because
the rates need to be quite suppressed at $T\simeq M_1$ in order not to
erase any lepton asymmetry generated during the decay, and
consequently are also suppressed with respect to the expansion rate at
higher temperatures. Hence,
additional interactions of the heavy neutrinos are required for them to
have a thermal distribution prior to the decay, so that the
leptogenesis scenario can be successful. This point was recently remarked in
ref.~\cite{pl96}, where it was shown that gauge interactions which are
naturally present in GUT scenarios (new $Z'$ or SU(2)$_R$ gauge
bosons) can easily do the job of producing a thermal population of
heavy neutrinos at $T>M_1$. 

Turning now to the supersymmetric version of this scenario, the
interactions of the heavy (s)neutrino field can be derived from the
superpotential
\beq
W=\frac{1}{2}M_i N_i
N_i+\lambda_{ij}\epsilon_{\alpha\beta}L_i^\alpha H^\beta N_j .
\eeq
Supersymmetry breaking terms are of no relevance for the mechanism of
lepton number generation and we are then left with the following
trilinear couplings in the Lagrangian, in terms of four component spinors,
\beq
{\cal L}=-\lambda_{ij}\epsilon_{\alpha\beta}\left\{M_j\tilde
N^*_j\tilde L^\alpha_iH^\beta+\overline{N_j}P_L\ell_i^\alpha
H^\beta+\overline{(\tilde h^\beta)^c}P_L\ell_i^\alpha \tilde N_j+
\overline{(\tilde h^\beta)^c}P_LN_j \tilde L_i^\alpha
  \right\}+h.c.\ .
\eeq

\begin{figure}
\centerline{\epsfbox{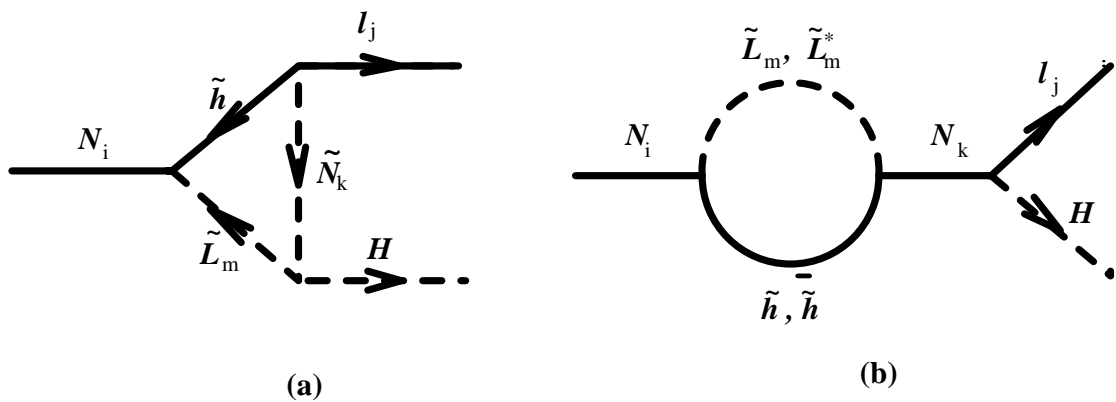}}
\caption{Supersymmetric contribution to the diagrams leading to 
CP violation in the $N\to\ell H$ decay.}
\end{figure}

{}From these couplings we obtain the tree level relations
\beq
\Gamma_{N_i\ell}+\Gamma_{N_i\bar\ell}=\Gamma_{N_i\tilde L}+\Gamma_{N_i\tilde
L^*}=\Gamma_{\tilde N^*\ell}=\Gamma_{\tilde N\tilde L}=
{(\lambda^\dagger\lambda)_{ii}
\over 8\pi}M_i .
\eeq

There are now new diagrams, like the ones in Figure~2,
 contributing to the generation of a leptonic
asymmetry. 
We will denote the corresponding asymmetry parameters 
 in the supersymmetric case with a tilde, so that for instance 
 $\tilde\epsilon^N_\ell(vertex)$ will receive contributions 
from the `standard' diagram in Figure 1a as well as from the one in
Figure~2a which involves superpartners in the loop\footnote{For later
convenience, we define $\tilde\epsilon^N_\ell$ as in Eq.~(2), rather
than normalizing it to the total $N$ decay rate which also includes
the slepton final states.}.
We also define the slepton asymmetry associated to $N$ decays as
\beq
\tilde\epsilon^N_{\tilde L}\equiv {\Gamma_{N\tilde L}-\Gamma_{N\tilde
L^*}\over \Gamma_{N\tilde L}+\Gamma_{N\tilde L^*}},
\eeq
which arises from diagrams similar to those in Figures~1 and 2 but
with $\tilde L \tilde h$ in the final state.
 We  similarly
define the asymmetry parameters associated to
sneutrino decays as
\beq
\tilde\epsilon^{\tilde N^*}_\ell\equiv {\Gamma_{\tilde
N^*\ell}-\Gamma_{\tilde N\bar\ell
}\over \Gamma_{\tilde N^*\ell}+\Gamma_{\tilde N\bar\ell}}  \ \ \ , \ \ \ 
\tilde\epsilon^{\tilde N}_{\tilde L}\equiv {\Gamma_{\tilde N\tilde
L}-\Gamma_{\tilde N^*\tilde L^*}\over \Gamma_{\tilde N\tilde
L}+\Gamma_{\tilde N^*\tilde L^*}}.
\eeq 

After a direct computation of these quantities we obtain
\beq
\tilde \epsilon^{N_i}_\ell(vertex)=-{1\over 8\pi}\sum_k 
g\left(y_k\right){\cal I}_{ki} \ ,
\eeq
where $g(x)= \sqrt{x}\  {\rm ln}\left[(1+x)/x\right]$, while
\beq
\tilde \epsilon^N_\ell(wave)=2\epsilon^N_\ell(wave).
\eeq

For the remaining channels we get similar results
\beq
\tilde \epsilon^N_{\tilde L}(vertex)=\tilde \epsilon^{\tilde
N^*}_\ell(vertex)=\tilde \epsilon^{\tilde N}_{\tilde L}(vertex)=
\tilde \epsilon^N_\ell(vertex),
\eeq
and also the wave function contributions are equal
\beq
\tilde \epsilon^N_{\tilde L}(wave)=\tilde \epsilon^{\tilde
N^*}_\ell(wave)= \tilde \epsilon^{\tilde N}_{\tilde L}(wave)=
\tilde \epsilon^N_\ell(wave).
\eeq

The vertex pieces were computed previously in ref.~\cite{ca93} and our
results are proportional. The sneutrino decay asymmetry was
also considered in ref.~\cite{mu93} in the limit in which all heavy
(s)neutrinos are degenerate and, specialised to that case, our results
are in agreement (except for the overall sign). 
The wave function contributions were not considered previously
and, as in the non--supersymmetric case, they are non--negligible. 
Let us also note that there are additional one loop diagrams involving 
the four--scalar couplings appearing in the $F$--terms of the scalar
potential. 
Although helpful to cure the divergences in the loops, they do not
contribute  to the asymmetry generation.

For the study of the decays of thermal populations of heavy neutrinos
and sneutrinos, 
it is convenient to introduce the total asymmetry defined as

\beq
\tilde\epsilon_i\equiv  \tilde\epsilon^{N_i}_\ell
+\tilde\epsilon^{N_i}_{\tilde L}+\tilde\epsilon^{\tilde
N_i^*}_\ell+\tilde\epsilon^{\tilde N_i}_{\tilde
L}=4\tilde\epsilon^N_\ell,
\eeq
resulting in
\beq
\tilde\epsilon_i=-\frac{1}{2\pi}\sum_{k\neq i}\left[
g(y_k)+{2\sqrt{y_k}\over y_k-1}\right]{\cal I}_{ki}.
\eeq

In particular, in the case of hierarchical masses ($y_k\gg 1$), 
we find that again the wave contribution becomes twice as large as the
vertex one, giving 
\beq
\tilde\epsilon_1\simeq -\frac{3}{2\pi}\sum_{k\neq 1}{1\over \sqrt{y_k}} 
\ {\cal I}_{k1}.
\eeq

With the definition in Eq.~(23), the resulting lepton asymmetry
is\footnote{Had we normalised $\epsilon^N_{\ell,\tilde L}$ 
to the total $N$ decay
rate, their contribution to $n_L$ would have to be multiplied by $g_N=2$
as in Eq.~(9).}
\beq
\frac{n_L}{s}=\frac{\tilde\epsilon_1}{s}\frac{T^3}{\pi^2}\simeq
1\times 10^{-3}\tilde\epsilon_1,
\eeq
where we have used that 
the effective number of degrees of freedom in the supersymmetric
scenario is approximately doubled, i.e. $g_*^{SUSY}\simeq
2\ g_*^{SM}$. Hence, to account for the baryon asymmetry of the Universe
we need now
\beq
\sum_{k\neq 1}{M_1\over M_k}\ {\cal I}_{k1}\simeq 0.7\times 10^{-6}.
\eeq
Comparing this with Eq.~(13), we see that the required parameters 
are similar in the supersymmetric and standard scenarios.

To summarize, we have computed all the contributions to the CP
violating asymmetries arising at one--loop in the decays of heavy
(s)neutrinos, both in the standard non--supersymmetric
Fukugita--Yanagida scenario and in its supersymmetric version. We
have discussed  the different results present in the
literature and showed that the contribution from wave function mixing
is relevant in the computation of the CP violating asymmetries.
 The baryon number generated in both scenarios was also
obtained.

\bigskip\bigskip
%\acknowledgments
It is a pleasure to thank A. Masiero for helpful suggestions and 
S. Bertolini, N. Paver and A. Smirnov for discussions.

\vfill\eject


\begin{thebibliography}{100}


\bibitem{fu86} M. Fukugita and T. Yanagida, Phys. Lett. {\bf B174}
(1986) 45.

\bibitem{sphal}N. S. Manton, Phys. Rev. {\bf D28} (1983) 2212.
V. A. Kuzmin, V. A. Rubakov and M. E. Shaposhnikov, Nucl. Phys. {\bf B155}
(1985) 36.

\bibitem{lu92} M. A. Luty, Phys. Rev. {\bf D45} (1992) 455.

\bibitem{pl96} M. Pl\"umacher, hep--ph/9604229.


\bibitem{ge95} G. Gelmini and E. Roulet, Rep. Prog. Phys. {\bf 58}
(1995) 1207.

\bibitem{ca93} B. A. Campbell, S. Davidson and K. A. Olive,
Nucl. Phys. {\bf B399} (1993) 111.

\bibitem{mu93}H. Murayama et al., Phys. Rev. Lett. {\bf 70} (1993)
1912.

\bibitem{ig79}A. Yu. Ignatiev, V. A. Kuzmin and M. E. Shaposhnikov,
JETP Lett. {\bf 30} (1979) 688.

\bibitem{bo91}F. J. Botella and J. Roldan, Phys. Rev. {\bf D44} (1991)
966. 

\bibitem{li93} J. Liu and G. Segr\`e, Phys. Rev. {\bf D48} (1993)
4609.

\bibitem{fu94} M. Fukugita and T. Yanagida, {\it Physics of
Neutrinos,} in: {\it Physics and Astrophysics of Neutrinos},
Eds. M. Fukugita and A. Suzuki, (Springer--Verlag, Tokyo, 1994).

\bibitem{li94} J. Liu and G. Segr\`e, Phys. Rev. {\bf D49} (1994)
1342.

\bibitem{ha90} J. A. Harvey and M. S. Turner, Phys. Rev. {\bf D42}
(1990) 3344.

\bibitem{co95} C. J. Copi, D. N. Schramm and M. S. Turner, Science
{\bf 267} (1995) 192.

\end{thebibliography}
\end{document}